\documentclass[journal,onecolumn]{IEEEtran}
\usepackage{epsfig}
\usepackage{graphicx}
\usepackage{graphics}
\usepackage{epstopdf} 
\usepackage{tikz}
\usetikzlibrary{calc,shapes.geometric}
\usetikzlibrary{arrows,snakes,backgrounds}

\ifCLASSINFOpdf
\else
\fi

\usepackage[cmex10]{amsmath}

\def\clock{\n=\time \divide\n 60
  \m=-\n \multiply\m 60 \advance\m \time
  \ifnum \n>12 \advance\n -12 \fi
   \number\n.\twodigits\m~\ampm\time}
\def\ampm#1{\ifnum #1< 720 am\else pm\fi}
\def\twodigits#1{\ifnum #1<10 0\fi \number#1}

\usepackage{epsfig}

\def\hyptest{\renewcommand{\arraystretch}{-0.7} 
\begin{array}{c}  
\mbox{\tiny{$H_{1}$}}  \\ \vspace{-0.5 mm}
>\\ 
<\\  
\mbox{\tiny{$H_{0}$}} 
\end{array}
}

\def\nexto{\kern -0.54em}

\def\prob{{\rm {I\ \nexto P}}}

\def\pfa{{\rm P_{FA}}}

\newcount\m \newcount\n
\def\clock{\n=\time \divide\n 60
  \m=-\n \multiply\m 60 \advance\m \time
  \ifnum \n>12 \advance\n -12 \fi
   \number\n.\twodigits\m~\ampm\time}
\def\ampm#1{\ifnum #1< 720 am\else pm\fi}
\def\twodigits#1{\ifnum #1<10 0\fi \number#1}

\begin{document}

\title{Multipulse Order Statistic Constant False Alarm Rate Detector in Pareto Background}
\author{Graham  V. Weinberg  \\ (Draft created at \clock)\\
 }
\maketitle

\markboth{Multipulse OS-CFAR \today}%
{}

\begin{abstract}
In a recent study, the extension of sliding window detectors from the single to multipulse case has been considered. This short note continues the analysis of such detectors, and specifies an order statistic variation. The probability of false alarm is derived in a useful compact mathematical expression.
\end{abstract}


\section{Introduction}
Recently, \cite{weinberg19} specified a multipulse constant false alarm rate (CFAR) detector, generalising that derived in \cite{mezache},
for the case of a geometic mean (GM) CFAR. Since multiplie pulse based detectors are of interest, it is useful to consider the case where the test cells are merged to a single measurement using an order statistic. This brief paper outlines how this can be done.

The context of interest is non-coherent detection under a Pareto Type I clutter model assumption. Sliding window detector development can be found in \cite{gandhi} and \cite{minkler}, while a comprehensive account in the scenario of interest is \cite{weinbergbook}. The latter only considered detector development with a single cell under test (CUT). To generalise this, \cite{mezache} considered the situation where multiple pulses are available, or equivalently, a series of CUTs. The authors examined GM detectors, as well as greatest of and smallest of decision rules, based upon the GM. However, these multipulse detectors were based upon the partial CFAR processes introduced in \cite{weinberg13}. A recent study showed how the full CFAR property could be
achieved, for the GM-based detector \cite{weinberg19}. This has utilised the developments in \cite{weinberg17}.

The decision rules of interest assume the existence of a series of non-negative measurements, denoted $Z_1, Z_2, \ldots, Z_N$, which are assumed to be independent and identically distributed. These measurements are referred to as the clutter range profile (CRP). The context for this work is X-band maritime surveillance radar; hence it will be assumed that these have a Pareto Type I distribution. Hence for all $j \in \{1, 2, \ldots, N\}$, 
\begin{equation}
F_{Z_j}(t) = \prob(Z_j \leq t) = 1 - \left( \frac{\beta}{t}\right)^\alpha, \label{parcdf}
\end{equation}
for $t\geq \beta$, and is zero otherwise. Here $\alpha>0$ is the shape and $\beta > 0$ is the scale parameter. This distribution function has a support not beginning at zero. The justification of a Pareto Type I model, for the scenario of interest, has been documented in \cite{weinbergbook}. To paraphrase the latter, the original fits to real X-band maritime surveillance radar clutter showed that a Pareto Type II model is appropriate. Since in many cases the Pareto scale parameter $\beta << 1$ it follows that the Pareto Type I model can be used as a basis for detector design.

In the single pulse case, a CUT is taken, which is assumed to be independent of the CRP, and denoted by $Z_0$. This is also a non-negative random variable. 
Sliding window detectors apply some function $f = f(Z_1, Z_2, \ldots, Z_N)$ to the CRP to produce a single measurement of the clutter level.
This is then normalised by a constant $\tau > 0$, called the threshold multiplier. Suppose that $H_0$ is the hypothesis that the CUT does not contain a target, while $H_1$ is the hypothesis that it contains a target embedded in clutter.
A typical test can be written
\begin{equation}
Z_0 \hyptest \tau f(Z_1, Z_2, \ldots, Z_N), \label{test1}
\end{equation}
where the notation in the above means that $H_0$ is rejected only if $Z_0 > \tau f(Z_1, Z_2, \ldots, Z_N)$. 

The Pfa of test \eqref{test1} is given by
\begin{equation}
\pfa = \prob(Z_0 > \tau f(Z_1, Z_2, \ldots, Z_N) | H_0). \label{pfa1}
\end{equation}
For a given Pfa and function $f$, one can solve for $\tau$ for application in \eqref{test1}. If this can be done in such a way that the Pfa does not vary with the clutter power, then the test is said to have the CFAR property. The importance of this property is evident from the fact that if there is variation with the resulting Pfa, this can cause series problems when the detector outputs are applied to a tracking algorithm, as an example. Hence sliding window detectors, with the CFAR property, are highly desirable.

When detectors of the form \eqref{test1} are applied in the Pareto clutter case, it is not possible to achieve the CFAR property completely. Hence, transformation approaches have been developed, and when coupled with complete sufficient statistics, it is possible to introduce variations of \eqref{test1} with the full CFAR property \cite{weinberg17}.

As an example of the above,  the decision rule
\begin{equation}
Z_0 \hyptest Z_{(1)}^{1-\tau} Z_{(k)}^\tau, \label{osdet1}
\end{equation} 
where $1 < k \leq N$ and $Z_{(k)}$ is the $k$th order statistic (OS) of the CRP, can be shown to have Pfa given by
\begin{equation}
\pfa = \frac{N!}{(N+1)(N-k)!} \frac{ \Gamma(\tau + N - k +1)}{\Gamma(\tau + N)}, \label{pfa1}
\end{equation}
when operating in Pareto Type I clutter, implying it is a complete CFAR detector. 

The next section extends \eqref{osdet1} to the multipulse situation.

\section{Multipulse OS-CFAR}
Here it is assumed that there are a series of $M$ pulses available, so that there are $M$ CUTs. Suppose that $X_{(k)}$ is the $k$th OS of the CUTs.
It is assumed that there is a single CRP available, consisting of $N$ statistics $Z_1, Z_2, \ldots, Z_N$, as before. Then a multipulse variation of \eqref{osdet1} is
\begin{equation}
X_{(k)} \hyptest  Z_{(1)}^{1-\tau} Z_{(n)}^\tau, \label{osdet2}
\end{equation}
where $1 \le k \leq M$, $1 < n \leq N$ and $Z_{(n)}$ is the $n$th OS for the CRP. This provides a multipulse OS-based alternative to the multipulse GM-CFAR derived in \cite{weinberg19}. The fact that \eqref{osdet2} is completely CFAR follows from the Pareto-exponential duality property \cite{weinbergbook}. To illustrate this, suppose that $Z$ is a random variable with distribution function \eqref{parcdf}, and suppose that $X$ is an exponential random variable with parameter unity. Then it can be shown that
\begin{equation}
Z = \beta e^{\alpha^{-1} X}, \label{parexp}
\end{equation}
and it is useful to observe that \eqref{parexp} extends to OS. 

Suppose that $X_j^*$ and $Z_j^*$ denote the duals of the CUT and CRP statistics respectively,  and that $X_{(j)}^*$ and $Z_{(j)}^*$ are the OS duals.
By an application of \eqref{parexp}, it can be shown that the Pfa of \eqref{osdet2} is
\begin{equation}
\pfa = \prob(X_{(k)}^* > (1-\tau)Z_{(1)}^* + \tau Z_{(n)}^*). \label{pfa2}
\end{equation}
Since the minimum $Z_{(1)}^*$ has an exponential distribution with parameter $N$, the Pfa can be written
\begin{equation}
\pfa = \int_0^\infty N e^{-Nt} \prob( X_{(k)}^* > t + \tau\left[ Z_{(n)}^* - t | Z_{(1)} = t \right] dt. \label{pfa3}
\end{equation}
It is shown in \cite{weinberg17} that  $Z_{(n)}^* - t | Z_{(1)} = t $ is the $(n-1)$th OS of a series of $N-1$ independent and identically distributed exponential random variables, with parameter unity. Hence if its density is denoted $f_{Q_{(n-1)}^*}$ then
\begin{eqnarray}
\pfa &=& \int_0^\infty \int_0^\infty Ne^{-Nt} f_{Q_{(n-1)}^*}(q) \prob(X_{(k)}^* > t + \tau q)dt dq \nonumber\\
\nonumber\\
&=& \int_0^\infty \int_0^\infty Ne^{-Nt} (n-1) {N-1 \choose n-1} (1-e^{-q})^{n-2} e^{-q(N-n+1)} \prob(X_{(k)}^* > t + \tau q)dt dq . 
\label{pfa4}
\end{eqnarray}
Now it can be shown that
\begin{equation}
 \prob(X_{(k)}^* < t + \tau q) = \sum_{i=k}^M {M \choose i} e^{-(t + \tau q)(M-i)} ( 1 - e^{-(t+\tau q)})^i, \label{pfa5}
\end{equation}
(see  \cite{weinbergbook}).
Hence, by an application of \eqref{pfa5} to \eqref{pfa4}, it follows that
\begin{eqnarray}
1  - \pfa &=& \sum_{i=k}^M {M \choose i} N(n-1) { N-1 \choose n-1} \times \nonumber\\
\nonumber\\
&& \int_0^\infty \int_0^\infty e^{-Nt} (1-e^{-q})^{n-2}  e^{-q(N-n+1)} e^{-(M-i)(t+\tau q)}
[ 1 - e^{-(t+\tau q)}]^i dt dq. \label{pfa6}
\end{eqnarray}
The double integral in \eqref{pfa6} can be reduced by a series of transformations. Firstly, by applying $\phi = e^{-t}$ and then $\psi = e^{-q}$ the double integral reduces to 
\begin{equation}
\int_0^1 \int_0^1 \phi^{N-n+\tau(M-i)} (1-\psi)^{n-2} \phi(N+M-i-1)(1-\phi \psi^\tau)^i d\phi d\psi. \label{pfa7}
\end{equation}
Then by application of the binomial theorem, 
\begin{equation}
(1-\phi  \psi^\tau)^i = \sum_{l=0}^i { i \choose l} (-1)^l \phi^l \psi^{\tau l}.\label{pfa8}
\end{equation}
Finally, by applying \eqref{pfa8} to \eqref{pfa7}, it becomes possible to evaluate the integral with respect to $\phi$ exactly, while the integral with respect to 
$\psi$ can be identified as a beta function. When the result is simplified, \eqref{pfa6}  becomes
\begin{eqnarray}
1 - \pfa &=& \frac{N!}{(N-n)!} \sum_{i=k}^M {M \choose i} \sum_{l=0}^i {i \choose l} (-1)^l \
 \frac{ \Gamma(N-n + \tau(M-i+l) + 1)}{ (N+M - i+l) \Gamma(N + \tau(M-i+l))}. \label{pfa9}
\end{eqnarray}
Hence, for given Pfa, $N$, $M$, $n$ and $k$, one can determine $\tau$ from \eqref{pfa9}, for application to \eqref{osdet2}, with numerical methods.

This detector, together with the GM counterpart in \cite{weinberg19}, will be investigated numerically in subsequent versions of this paper.


\begin{thebibliography}{}

\bibitem{weinberg19}
Weinberg, G. V. (2019). 
Extension of the Geometric Mean Constant False Alarm Rate Detector to Multiple Pulses.
ArXiv Preprint, https://arxiv.org/abs/1901.00250.

\bibitem{mezache}
Sahed, M., Mezache, A. (2015).
Analysis of CFAR Detection with Multiple Pulses Transmission Case in Pareto Distributed Clutter.
4th International Conference on Electrical Engineering, Algeria.


\bibitem{gandhi}
Gandhi, P. P., Kassam, S. A. (1988). 
Analysis of CFAR Processors in Nonhomogeneous Background. 
IEEE Transactions on Aerospace and Electronic Systems, 24,  427-445.


\bibitem{minkler}
Minkler, G.,  Minkler, J.  (1990). 
CFAR: The Principles of Automatic Radar Detection in Clutter. Magellan.


\bibitem{weinbergbook}
Weinberg, G. V. (2017).
Radar Detection Theory of Sliding Window Processes. CRC Press.

\bibitem{weinberg13}
Weinberg, G. V. (2013).
Constant False Alarm Rate Detectors for Pareto Clutter Models.
IET Radar, Sonar and Navigation, 7, 153-163.


\bibitem{weinberg17}
Weinberg, G. V. (2017).
On the Construction of CFAR Decision Rules via Transformations.
IEEE Transactions on Geoscience and Remote Sensing, 55, 1140-1146.



\end{thebibliography}
\end{document}